# An Empirical Investigation on the Challenges of Creating Custom Static Analysis Rules for Defect Localization

Diogo S. Mendonça[1, *], Marcos Kalinowski [2]

[1]*Centro Federal de Educação Tecnológica Celso Suckow da Fonseca (CEFET/RJ), Rio de Janeiro, Brazil*
*diogo.mendonca@cefet-rj.br*

[2]*Pontifícia Universidade Católica do Rio de Janeiro (PUC-Rio), Rio de Janeiro, Brazil*
*kalinowski@inf.puc-rio.br*

**Abstract:**

**Background**: Custom static analysis rules, i.e., rules specific for one or more applications, have been successfully applied to perform corrective and preventive software maintenance. Pattern-Driven Maintenance (PDM) is a method designed to support the creation of such rules during software maintenance. However, as PDM was recently proposed, few maintainers have reported on its usage. Hence, the challenges and skills needed to apply PDM properly are unknown. **Aims:** In this paper, we investigate the challenges faced by maintainers on applying PDM for creating custom static analysis rules for defect localization. **Method:** We conducted an observational study on novice maintainers creating custom static analysis rules by applying PDM. The study was divided into three tasks: (i) identifying a defect pattern, (ii) programming a static analysis rule to locate instances of the pattern, and (iii) verifying the located instances. We analyzed the efficiency and acceptance of maintainers on applying PDM and their comments on task challenges. **Results:** We observed that previous knowledge on debugging, the subject software, and related technologies influenced the performance of maintainers as well as the time to learn the technology involved in rule programming. **Conclusions:** The results strengthen our confidence that PDM can help maintainers in producing custom static analysis rules for locating defects. However, a proper selection and training of maintainers is needed to apply PDM effectively. Also, using a higher level of abstraction can ease static analysis rule programming for novice maintainers.

* Corresponding author.

# 1 Introduction

Maintenance is the most costly phase in the software lifecycle (Bourque et al., 2014). Defect prevention and correction activities consume part of these resources. Additionally, the impact of failures in software in use can have a wide range of variation, going from slight inconvenience to severe damage, including economic ones (Jones & Bonsignour, 2011).

A way to reduce costs of preventive and corrective maintenance is to automate the localization of source code defects. Static analysis is a commonly used approach for alerting defects (Heckman & Williams, 2011; Muske & Serebrenik, 2016). Some tools, such as SonarQube (SonarSource, 2008) and PMD (InfoEther Inc, 2020), support alerting general purpose code defects and the development of custom rules to locate application-specific ones.

In a survey with developers (Tymchuk et al., 2018), one third of them agreed that good (useful) rules are related to a specific context, such as a particular project. Although custom rules are useful and commonly available, they are not well explored in practice (Beller et al., 2016; Christakis & Bird, 2016). Less than 5% of the static analysis rules used in open source projects are custom rules (Beller et al., 2016). Furthermore, only 8% of developers reported using custom rules in practice (Christakis & Bird, 2016).

The Pattern-Driven Maintenance (PDM) method (Mendonça et al., 2018) was designed for helping maintainers on automating the localization of defects during maintenance. In this method, the maintainer performs a systematic set of steps to identify a defect pattern, produce custom static analysis rules for locating defect instances, and evaluate the rules produced during software maintenance. PDM already helped to improve the reliability and security of industrial applications (Mendonça et al., 2018). Furthermore, the rules produced by PDM showed to be reusable in other applications that have similar technologies and programming style (Mendonça & Kalinowski, 2020). However, as PDM was recently created (Mendonça et al., 2018), few maintainers have reported its usage. Hence, the challenges, the skills needed, and the acceptance of PDM by maintainers are unknown.

In this paper, we investigate the challenges faced by maintainers on applying PDM for creating custom static analysis rules for defect localization. We aim to answer the following research questions:

*RQ1. What are the challenges faced by maintainers while applying PDM for creating custom static analysis rules?*

*RQ2. Would maintainers accept to use PDM?*

We conducted an observational study on novice maintainers creating custom static analysis rules by applying PDM to answer these research questions. The study was divided into three tasks: (i) identifying defect patterns, (ii) programming a static analysis rule to locate instances of a pattern, and (iii) verifying the located instances. We analyzed the efficiency of maintainers on applying each task and their comments on task-related challenges. We also analyzed the acceptance of PDM by the maintainers by applying a Technology Acceptance Model (TAM) based questionnaire (Davis, 1989).

We found that maintainers had difficulties in applying PDM. Few of them correctly completed the tasks. We observed that previous knowledge on debugging, the subject software, and related technologies had an influence on maintainers' performance. For the static analysis rule programming task, we first assumed that the maintainers would be able to implement the rule directly using an the source code Abstract Syntax Tree (AST). However, in a first trial, we observed that, even with a short training session, none of them was able to correctly implement the rule. Hence, we concluded that expecting developers which are not rule experts to develop defect rules directly using the AST was unrealistic. Therefore, we redesigned the second task to use a Domain Specific Language (DSL) with abstractions to ease the rule programming (Crispe & Mendonça, 2021) and conducted two additional trials to better investigate the rule programming feasibility. In this new investigation, we observed that some (25%) of the maintainers were able to correctly implement the rule, but that time to learn the DSL influenced the effectiveness of rule programming. According to the TAM-based questionnaire, most maintainers found PDM useful but not easy to apply.

Hence, while PDM can help to improve application reliability and security (Mendonça et al., 2018), there is a clear need for training in order to apply it effectively. We provide guidance on which skills are needed to apply the method to produce custom static analysis rules, allowing practitioners to better select or train professionals to apply PDM.

The remainder of this paper is organized as follows. Section 2 describes the PDM method as background and related work. Section 3 outlines the observational study design on the challenges faced by maintainers and their acceptance of PDM. Section 4 presents the study results. Section 5 discusses the research questions based on the obtained results. Section 6 presents the threats to validity. Finally, Section 7 concludes the paper and presents future work.

## 2 Background and Related Work

In this section, we briefly explain the Pattern-Driven Maintenance (PDM) method (Mendonça et al., 2018) with a comprehensive example of its application and produced custom static analysis rules. We also present related work.

### 2.1. Pattern-Driven Maintenance

Figure 1 shows the activities collapsed into steps along with the control flow of the method. Two primary paths can be observed: the maintenance path (steps 1, 2, and 3) and the rule improvement cycle (steps 4, 5, and 2).

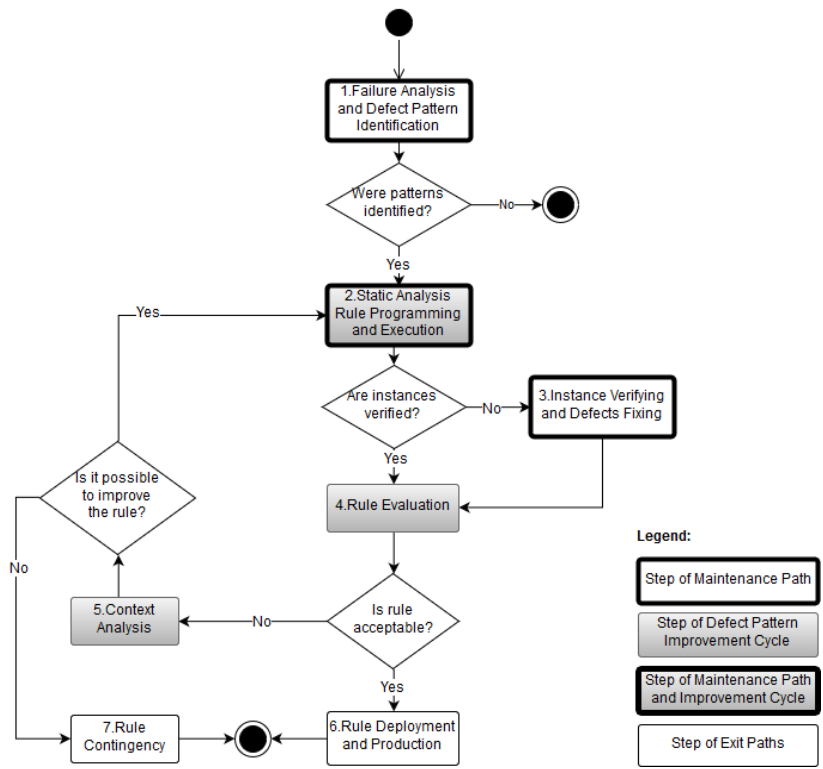

**Figure 1: The PDM method control flow**

The maintenance path includes activities to inspect or process failure data and identify defect patterns (step 1), to develop custom static analysis rules to locate the latent defects (step 2), and to verify the detected instances and correct the defects (step 3). The execution of the maintenance path occurs when there are new failures data. The failure data source must be monitored periodically to identify those new records by performing the first step of the method (failure analysis and defect pattern identification). Example of failures data sources are logs or issue reports. Eventually, no defect pattern will be identified in step 1, and in this case, no further step of PDM needs to be performed. The maintainer should perform the typical corrective maintenance in cases there are failures, but no patterns were identified. For simplicity, we did not represent this case in PDM workflow (Figure 1).

The rule improvement cycle is performed when the evaluation of the rules (step 4) (e.g., based on precision and recall) does not reach acceptable levels to alert during development. These levels vary according to the static analysis rule and depend on factors such as the impact on software reliability. Each company or maintenance team also has its own tolerance levels to false positive and negative alerts. Thus, we do not prescribe the thresholds for these levels. Further information on how we established those levels in our industrial evaluations and benchmarks are provided in Section 2.2.

When precision or recall levels are not acceptable, the source code context of the detected defects is analyzed to improve the static analysis rules (step 5). Those contexts represent specific cases when a defect alerted is a false positive, thus being necessary its removal from the rule. Finally, there are two exit steps in the exit path of the method – rule deployment for defect alerting (step 6) and rule contingency (step 7) –, which include using the rules and patterns only in a limited way. Further details on the seven depicted steps can be found in our previous works (Mendonça

& Kalinowski, 2020; Mendonça et al., 2018). A complete example of how to apply PDM is presented in the next section.

## 2.2. Example of PDM Application

In this section, we present a step by step example of the PDM method application in a comprehensive and straightforward case. We applied PDM in an open-source JEE software named Employment and Internship Management System[1] (SisGEE). Students developed this software during an undergraduate course at CEFET/RJ, and CEFET/RJ employees use it. The selected example is the same one used during the observational study described in Section 3.

The inputs for the PDM method application are the logs of SisGEE and a specific version of SisGEE[2] system that generated those logs. The first PDM step is failure analysis and defect pattern identification. Performing failure analysis, we extracted data from the logs in order to compare the failures against each other and search for defect patterns. Table 1 presents data extracted from the logs. We can observe that exception type and error message are equal for failures 1 and 3, as well as for failures 4 and 5. Similar failures should have the related source code inspected together for defect pattern identification. Figure 2 and Figure 3 show the source code that originated failure 1 and 3, respectively. We can observe that both exceptions were thrown by the *Integer.parseInt* method. A simple solution to handle those failures is surrounding *Integer.parseInt* method call with a try/catch construct. In this way, we documented the pattern identified and the proposed solution to support static analysis rule programming. Table 2 presents the documentation of the defect pattern identified.

**Table 1: Data extracted from logs during failure analysis**

| #Failure | File Name | Line | Exception Type | Error Message |
|----------|-----------|------|----------------|---------------|
| 1 | BuscaTermoAditivoServlet | 49 | java.lang.NumberFormatException | For input string: "" |
| 2 | IncluirTermoEstagioServlet | 60 | java.lang.ClassCastException | java.lang.Double cannot be cast to java.lang.Float |
| 3 | VisualizarTermoEAditivo | 43 | java.lang.NumberFormatException | For input string: "" |
| 4 | RenovarConvenioServlet | 44 | java.lang.NullPointerException | |
| 5 | VisualizarTermoEAditivo | 50 | java.lang.NullPointerException | |
| 6 | index.jsp | 4 | org.apache.jasper.JasperException | File [import_head.jspf] not found |

---

[1] https://github.com/diogosmendonca/sisgee
[2] https://github.com/diogosmendonca/sisgee/tree/d06207f

```
39  protected void doPost(HttpServletRequest request, HttpServletResponse response) throws ServletException, IOException {
40      Locale locale = ServletUtils.getLocale(request);
41      ResourceBundle messages = ResourceBundle.getBundle("Messages", locale);
42
43      String msg = null;
44      String idAluno = request.getParameter("idAluno");
45      String mat = request.getParameter("matricula");
46
47      Integer id = null;
48      System.out.println("Aqui >>> "+idAluno);
49      if(Integer.parseInt(idAluno)!=-1){
50      msg = ValidaUtils.validaObrigatorio("Aluno", idAluno);
51      if(msg.trim().isEmpty()) {
```

**Figure 2: BuscaTermoAditivoServlet source code near line 49**

```
32  protected void doGet(HttpServletRequest req, HttpServletResponse resp) throws ServletException, IOException {
33
34      String ide = req.getParameter("ide");
35      String ida = req.getParameter("ida");
36      String matricula = req.getParameter("matricula");
37      UF[] uf = UF.asList();
38      TermoEstagio termoEstagio=null;
39      TermoAditivo termoAditivo=null;
40
41      Aluno aluno=AlunoServices.buscarAlunoByMatricula(matricula);
42      if(ide!=null)
43      termoEstagio=TermoEstagioServices.buscarTermoEstagio(Integer.parseInt(ide));
44      if(ida!=null)
45      termoAditivo=TermoAditivoServices.buscarTermoAditivo(Integer.parseInt(ida));
```

**Figure 3: VisualizarTermoEAditivo source code near line 43**

**Table 2: Documentation of the defect pattern identified**

| | |
|---|---|
| **Defect Name** | Unchecked Integer |
| **Description** | The application throws an exception when a string parameter is parsed to an Integer. |
| **Exception Type and Failure Message** | java.lang.NumberFormatException, For input string: "<value>" |
| **Parameters in Failure Message** | <value> - the value of the parameter passed |
| **Example of Failure Message** | java.lang.NumberFormatException, For input string: "" |
| **Class and Method of Throw** | Integer, method parseInt |
| **Defect Characterization** | A call to the parseInt method not surrounded by a try/catch |
| **Defect Code Example** | String intParam = request.getParameter("intParam"); ... //unchecked exception Integer intValue = Integer.parseInt(intParam); |
| **Fixed Code Example** | Integer intValue = null; try{ intValue = Integer.parseInt(intParam); }catch(NumberFormatException e){ //handle the exception } |

The next step of PDM is static analysis rule programming. Using the documentation of the defect pattern, the maintainer uses some static analysis tool to implement a rule that locates the defects that match the defect pattern. In our example, we used Souce Code Pattern Language (SCPL)(Crispe & Mendonça, 2021), a language that uses markups embedded in Java code pattern examples to produce custom rules, for this task. Figure 4 presents a simplified version of the static analysis rule for the defect pattern documented in Table 2. This rule locates all *Integer.parseInt* function calls in any methods (//inAnyMethod markup identified at the pattern header) that are not surrounded by a try/catch (//not_exists markup before the try/catch statement). This method checks whether the name of the class is *Integer* and the method name is *parseInt*. After locating one instance of the *parseInt* method, *SCPL* checks its parents recursively in search of a try block, returning if the method is not inside a try block. Further information on how to program custom static analysis rules using SCPL can be found in (Crispe & Mendonça, 2021).

After programming a static analysis rule, maintainers should use it to locate defect candidates. Table 3 presents the alerts produced by running the static analysis rule, together with the results of the verification of the alerts (achieved by inspecting the source code). Table 4 presents additional potential defect candidates, located using an IDE search, which were not alerted by the static analysis rule. It is noteworthy that none of them concerned true defects (i.e., the rule indeed should not have alerted them). We calculated the precision and relative recall (Mendonça et al., 2018), resulting in a precision of 23% and a relative recall of 100%. The precision indicates the ratio of alerts that correspond to defects (i.e., the ratio of true positives). The relative recall is the ratio of defects alerted out of the total number of defects matching the intended defect pattern in the source code. Hence, the relative recall is a relative number, verified by conducting a broad IDE search and verifying the retrieved instances through manual inspection. The intention of the broad IDE search was to include all possible locations of the intended pattern but not all code of the software, making manual inspection feasible.

**Table 3: Alerts produced by the first version of the static analysis rule with the results of their verification**

| #Alert | File | Line | Result of Verification |
|---|---|---|---|
| 1 | BuscaTermoAditivoServlet.java | 49 | Defect (True Positive) |
| 2 | BuscaTermoAditivoServlet.java | 54 | No Defect (False Positive) |
| 3 | FormTermoAditivoServlet.java | 115 | No Defect (False Positive) |
| 4 | FormTermoAditivoServlet.java | 225 | No Defect (False Positive) |
| 5 | FormTermoAditivoServlet.java | 265 | No Defect (False Positive) |
| 6 | VerTermoAditivoServlet.java | 48 | No Defect (False Positive) |
| 7 | VisualizarTermoEAditivo.java | 43 | Defect (True Positive) |
| 8 | VisualizarTermoEAditivo.java | 45 | Defect (True Positive) |
| 9 | FormTermoEstagioServlet.java | 214 | No Defect (False Positive) |
| 10 | FormTermoEstagioServlet.java | 600 | No Defect (False Positive) |
| 11 | FormTermoEstagioServlet.java | 646 | No Defect (False Positive) |
| 12 | FormTermoEstagioServlet.java | 749 | No Defect (False Positive) |
| 13 | FormTermoRescisaoServlet.java | 81 | No Defect (False Positive) |

**Table 4: Defect candidates not alerted by the first version of the static analysis rule**

| #Alert | File | Line | Result of Verification |
|---|---|---|---|
| 1 | PrincipalTermo.java | 378 | No Defect (True Negative) |
| 2 | ValidaUtils.java | 232 | No Defect (True Negative) |
| 3 | ValidaUtils.java | 233 | No Defect (True Negative) |

The 23% precision of the rule is unacceptable. Hence, we started a defect pattern improvement cycle by conducting context analysis. In the context analysis, we inspect false positive alerts searching for fixing alternatives different from the ones already included in the defect pattern. A fixing alternative which is not included in the pattern generates false positives. Consequently, the inclusion of a new fixing alternative in the pattern can improve its precision. Table 5 presents a fixing alternative found during context analysis. After documenting the fixing alternative, we modified the static analysis rule to include it and executed the new version of the rule. The modified version of the rule eliminated all except of one false-positive alert (Alert #12) of Table 3. In this way, the new precision of the rule is 75%, and the relative recall remains 100%. Alert #12 has a different fixing alternative from the others, but the effort to include it in the defect pattern was not worthwhile to eliminate one single instance. The new rule version was accepted and deployed into a production environment for alerting developers.

```
1  //inAnyMethod
2  //not_exists
3  try{
4      //Alert: Surround Integer.parseInt with a try/catch block
5      Integer.parseInt(any);
6  }catch(AnyException any){}
```

**Figure 4: Example of static analysis rule implemented using SCPL**

**Table 5: Documentation of a fixing alternative found during context analysis**

| Defect Pattern | Unchecked Integer |
|---|---|
| Context Name | Inside Integer Validation |
| Context Description | It is not possible to throw an exception in *Integer.parseInt* call when the call is inside an if block that checks its parameter for integer format |
| Context Cause | Control flow avoids exception throw |
| Context Characterization | An *Integer.parseInt* call inside an if block that uses the result of *ValidaUtils.validInteger* in its expression and *ValidaUtils.validInteger was* called with the same parameter of *Integer.parseInt*. |
| Code Example | campo = "Aluno"; idAlunoMsg = ValidaUtils.validInteger(campo, idAluno); if (!idAlunoMsg.trim().isEmpty()) { Integer idAlunoInt = Integer.parseInt(idAluno); |

---

                    ...
                    }

---

## 2.3. Related Work

In a survey with developers, one third of them agreed that good (useful) rules are related to a specific context, such as a particular project (Tymchuk et al., 2018). Although custom rules are useful and commonly available, they are not well explored in practice (Beller et al., 2016; Christakis & Bird, 2016). Less than 5% of the static analysis rules used in open source projects are custom rules (Beller et al., 2016). Furthermore, only 8% of developers reported using custom rules in practice (Christakis & Bird, 2016).

Possible reasons for the low adoption of custom rules could be the learning curve and background needed to produce them (Christakis & Bird, 2016), the completeness of static analysis tools, and the unwillingness to create custom rules [8]. However, developers demonstrated a willingness to improve static analysis results by providing information to static analysis tools in their code (Christakis & Bird, 2016). To the best of our knowledge, our research is the first to provide empirical evidence regarding the challenges developers face when producing custom static analysis rules by applying a systematic method. In case, the method applied was PDM (Mendonça et al., 2018).

PDM was designed and initially applied to two small-sized industrial web applications (Mendonça et al., 2018), one implemented in Python/Django and the other in PHP. In both evaluations, given the observed failures, PDM application was able to eliminate all defects that could generate failures with similar causes, helping to improve the overall application reliability and security.

Although PDM allowed successfully improving the precision of the static analysis rules without harming the relative recall, the rules did not reach the desired precision levels of 80% established by the company in the first evaluation (it reached 68%). In the second evaluation, as SonarQube for PHP code supports data and control flow analysis features, it was possible to achieve a superior precision of 89.5-100%, being accepted by the company for defect prevention.

It was also observed that the way how PDM steps are performed impacts the application effort (Mendonça et al., 2018). Indeed, the PDM variation applied in the second evaluation, considering the context of false positives as soon as possible, showed to reduce the method application effort. Another factor that may have an impact on effort is the familiarity of the maintainer with the subject application. Finally, regarding the precision, the findings indicate that there is an influence of the technology selection on the precision of the rules. Using data and control flow analysis features helped to improve the precision of the rules in the second application.

Subsequently, we evaluated the reuse of rules produced in previous PDM applications (Mendonça & Kalinowski, 2020). The rules produced by applying PDM were reused in within- or cross-company environments. It was possible to find defects in other software by reusing rules, as well as to reduce the verification effort of defect patterns. Nevertheless, as expected, the architecture and programming style played an essential role in successfully reusing rules produced by PDM application, thus being an influencing factor for reuse (Mendonça & Kalinowski, 2020).

This finding indicates the feasibility of PDM producing rules that are application architecture and coding style specific, and not only application-specific. In this way, the reuse of rules has the advantage of producing more robust rules and might reduce the effort of identifying similar patterns in other systems. We also observed that previous successful experience with PDM influenced the reusable rules' adoption (Mendonça & Kalinowski, 2020). As a result of this investigation, we provided a set of recommended practices for the evaluation and implementation of reusable rules produced by PDM (Mendonça & Kalinowski, 2020).

## 3 Observational Study Design

Previous studies indicated that applying PDM is effective (Mendonça et al., 2018) for corrective and preventive maintenance, and that custom static analysis rules produced can be reused in other applications (Mendonça & Kalinowski, 2020). However, few maintainers applied the method. Nevertheless, their feedback about the PDM experience was positive, strengthening our confidence that PDM could help other maintainers. However, this feedback is insufficient to know if they would be able to effectively apply PDM and accept the technology, as well as the challenges faced. Aligned with the methodology for introducing software processes described by Shull *et al.* (2001), given that at this point the feasibility of PDM was already determined, the next step was to conduct an observational study.

In this way, our research objective is to evaluate PDM concerning the effectiveness and acceptance from the viewpoint of different maintainers, answering the following research questions:

*RQ1. What are the challenges faced by maintainers while applying PDM for creating custom static analysis rules?*

*RQ2. Would maintainers accept to use PDM?*

We conducted an observational study of PDM application using five groups of novice maintainers with different experiences and knowledge. The first group served as a pilot for instrument validation and was composed of computer science graduate students (n=9). The other four groups were composed of computer science undergraduate students. Students from group A (n=27), C (n=18), and D (n=10) had no previous experience with the software under investigation and limited experience with the involved technologies (JEE), whereas students from group B (n=18) had previous experience with the software and were more familiar with its technologies.

Groups A and B were trained and applied the two main reasoning steps involved in PDM, concerning identifying defect patterns and evaluating them for improvement. Additionally, for group B we conducted the static analysis rule programming task to investigate whether maintainers would be able to implement the rule directly using an the source code Abstract Syntax Tree (AST). Unfortunately, even with a short training session, none of them was able to correctly implement the rule. We concluded that expecting developers which are not rule experts to develop defect rules directly using the AST was unrealistic. Therefore, we redesigned the second task to use a Domain Specific Language (DSL) with abstractions to ease the rule programming (Crispe & Mendonça, 2021) and conducted new trials with groups C and D to further investigate the rule programming feasibility.

We collected the results of applying those tasks and their feedback on the difficulties found. We also used the Technology Acceptance Model (TAM) (Davis, 1989) to assess the acceptance of PDM by maintainers in its three dimensions: ease of use, usefulness, and intention of use.

The remainder of this section is organized based on the planning part of a guideline for reporting experiments described by Jedlitschka *et al.* (2008).

### 3.1.Goal

The research objective covered in this study is to evaluate PDM concerning the effectiveness of maintainers and the acceptance and from the viewpoint. In this way, following the GQM template (Basili et al., 1994) we have the following goal:

*Analyze* PDM *for the purpose of* characterization *with respect to* challenges faced on conducting its steps, perceived usefulness, ease of use, and intention of use *from the point of view of* maintainers *in the context of* computer science students applying the PDM steps on excerpts of artifacts from a real and specific software product.

### 3.2. Participants

We selected the subjects of the study by convenience. We had access to graduate and undergraduate students in courses related to software quality of two different Brazilian universities. The first group of students was composed of nine graduate students in informatics from the Pontifical Catholic University of Rio de Janeiro (PUC-Rio). We called this group Pilot since its main purpose was helping to validate our materials. The other four groups, A (n=27), B (n=18), C (n=18), and D (n=10) were composed of undergraduate students in computer science from PUC-Rio (A and C) and from the Federal Center of Technological Education Celso Suckow da Fonseca (CEFET/RJ) (B and D). Students of Group A and C were enrolled in a discipline on software testing and measurement, which is in the second year of their course, whereas students of Group B and D were enrolled in a discipline on software engineering, which is in the third year of their course.

One relevant difference between the groups was that group B had previous experience with the software on which they would apply PDM. The previous experience was possible because that software was used in the final discipline assignment in which group B students were enrolled. At the time when the students performed the tasks of the study, the assignment was already passed to the students.

### 3.3. Experimental Materials

All experiment materials are available online in our replication package[3] in the zenodo.org open science repository. A description of these materials is provided hereafter.

The characterization of students was made by filling a characterization form with questions about their experience (in months) with software development and maintenance in different contexts (for their own use, in a course, and in the industry). We also included questions about the level of experience with techniques and technologies that could influence the results of the experiments. In this case, we asked about their level of experience in Java, JEE, stack trace reading, static analysis rule programming, and source code inspections, as well as their proficiency in the English language.

---

[3] http://doi.org/10.5281/zenodo.5552623

The software for applying PDM was selected by convenience. The selected Internship and Employment Management System (SisGEE)[4] is an information system developed by students as an assignment of a web programming discipline of a computer science course at CEFET/RJ. SisGEE was developed using JEE technology and contained some defect patterns of unhandled latent exceptions in its source code.

We exercised some of those unhandled exceptions to produce a log for PDM application (see Table 1). The version of SisGEE that was used for producing this log is available on Github[5]. The log contains two failures produced for invalid conversion from string to an integer (NumberFormatException), two failures produced by access in a service layer that returns null, and the null value is used without previously checking (NullPointerException), and other two failures that do not form any pattern.

The first task of the study (Task 1) consists of executing the first PDM step, i.e., failure analysis and defect pattern identification. The failure identification consists of extracting failure data from logs filling a provided form. The groups of maintainers that participated in this task (Pilot, A, and B) received the same form for failure identification. The data that should be extracted consists of a file name and line where the exception was thrown as well as the exception type and error message contained in the failure. After performing failure identification, maintainers were instructed to use the extracted data to compare failures and identify similar ones.

Thereafter, maintainers were instructed to inspect the source code related to similar failures to identify patterns formed by the defects. If a defect pattern was identified, maintainers should document it. The Pilot group received a form with separate fields for information that would be useful for identifying a defect pattern, whereas groups A and B received training in using a pattern language and should document the defect patterns using that language. Table 2 presents an example of this form filled. The pattern language used by groups A and B consists of the same syntax of the software programming language (Java) but including wildcard symbols and conventions for documenting the pattern. Table 6 presents the wildcards and conventions available in the pattern language while Table 7 presents an example of defect pattern documented using this language.

**Table 6: Pattern language wildcards and conventions**

| Description | Wildcard Symbol | Example |
| --- | --- | --- |
| An element must be present as is it appear in the defect instances to describe the pattern | Use the same elements present in the examples of the defect, typically structural element. | If, for, while, switch, assignments, operators, etc. |

---

| An element must be present to describe the pattern, but the identifier name can variate in each instance of the defect. An abstraction of identifier name is needed. | Prefix name convention: any, some, other; followed by the name of the element needed. | anyVariable, otherVariable, someMethod, someClass |
| --- | --- | --- |
| An element or group of elements do not need to be present to form a pattern, and it can be fully abstracted in the pattern description | The symbol "…" is used where any code can be present. | if(someVariable){ <br> … <br> }, <br> someClass.someMethod(…) |

**Table 7: Example of defect pattern documented using pattern language**

| Defect Examples | Defect Pattern |
| --- | --- |
| ```String param = request.getParameter("param"); … if(param.length() > 0){ /* line where the exception was thrown */ … }``` | ```String someVariable = request.getParameter("someParam"); … someVariable.anyMethod(...); …``` |
| ```String other = request.getParameter("otherParam"); … String msg = "invalid value: " + other.trim(); /*line where the exception was thrown */ …``` | |

After performing each task of the study, the maintainers were asked to fill a follow-up questionnaire with questions about their strategies and perceptions on the task. The questionnaire used for the first task of the study was equal for all groups of maintainers. The questions asked concerned: the strategy used by the maintainer to identify the defect pattern, the perception if the time was enough to complete the task, the confidence in the patterns reported, the ease of performing the task, and the difficulties found.

Task 2 consisted of programming a static analysis rule. For this task, one defect pattern documentation was provided and the maintainers were asked to program a static analysis rule that locates the instances of this defect pattern. The provided defect pattern documentation was the one presented in Table 2, which concerns a NumberFormatException. We understand that such exception could be handled in caller methods/functions. However, the intention of the study was to evaluate the factors of influence on developing simple rules. Thus, the pattern described in Table 2 to be implemented considered only exception handling within the same method.

The groups that participated in Task 2 were group B, C, and D. For group B we asked participants to implement the rule directly in Java using the source code abstract syntax tree and the Visitor design pattern (Gamma et al., 1995), an approach that is commonly applicable in static analysis tool. Nevertheless, none of them was able to correctly implement the rule. Therefore, we redesigned Task 2 to use a Domain Specific Language (DSL) with abstractions to ease the rule programming for the trials conducted with groups C and D. The tool selected for static analysis rule programming was Source Code Pattern Language (SCPL) (Crispe & Mendonça, 2021), which supports implementing rules for Java programs using markups in code pattern examples. After finishing the task, the maintainers were asked to provide the source code of the programmed static analysis rule and fill the follow-up questionnaire, which follows the same template of Task 1.

Finally, Task 3 comprised the PDM steps of rule evaluation and context analysis. To perform this task, we provided the documentation of one defect pattern, the source code of the application that contains this defect pattern, and a list of source code lines in this application that were alerted by a static analysis rule that implements the defect pattern. Table 2 presents the provided defect pattern documentation. The application source code was the same one of the other tasks. The alerted source code lines were provided to the maintainers in a form.

During Task 3, maintainers should classify the alerts provided as defects or false positives. If a false positive was found, they should inform which fixing alternative was present in the source code. In the case of finding new fixing alternatives, maintainers should document them. The pilot group documented the fixing alternatives using a form while groups A and B used the pattern language. After performing the task, the maintainers were asked to fill the follow-up questionnaire, which is similar to the follow-up questionnaire of other tasks.

At the end of the study, the maintainers were asked to fill the TAM questionnaire. This questionnaire is composed of nine questions split into three dimensions: usefulness, ease of use, and intention to use. The answers are provided in a five-point Likert scale ranging from strongly disagree to strongly agree. The questions of the TAM questionnaire, adjusted to our study, are presented in Table 8.

**Table 8: TAM questions used in the study**

| Dimension | ID | Question |
|-----------|-----|----------|
| Usefulness | Q1 | Using *PDM* would improve my performance in preventing defects (i.e.., prevent faster)? |
| | Q2 | Using *PDM* would improve my productivity in preventing defects (i.e.., prevent more and faster)? |
| | Q3 | Using *PDM* would enhance my effectiveness in preventing defects (i.e., prevent more)? |
| | Q4 | I would find *PDM* useful in preventing defects? |
| Ease of use | Q5 | Learning to operate *PDM* would be easy for me? |
| | Q6 | I would find it easy to get *PDM* to prevent defects? |
| | Q7 | It would be easy for me to become skillful in the use of *PDM*? |
| | Q8 | I would find *PDM* easy to use? |
| Intention to use | Q9 | I intend to use *PDM* regularly at work? |

### 3.4. Question Refinement and Variables

The first knowledge question we wanted to answer (RQ1) concerned the challenges faced by maintainers, i.e., for each task of the study, we want to understand how effective maintainers are in performing the task and the difficulties found. We used the percentage of maintainers that completed each task with success to provide us insights about how easy it is for a maintainer to effectively perform the task. Having this in mind, we aim at answering the following more detailed questions regarding the effectiveness:

RQ1. What are the challenges faced by maintainers while applying PDM for creating custom static analysis rules?
    a.  What are the challenges faced by maintainers while identifying and documenting defect patterns?
    b.  What are the challenges faced by maintainers while programming a custom static analysis rule?
    c.  What are the challenges faced by maintainers while identifying and documenting fixing alternatives present in false positives of a rule?

Some knowledge and experiences might influence applying PDM. Hence, we were interested in having insights on how these affect the effectiveness of maintainers in applying PDM. Therefore, in our study, we additionally investigated if knowledge and experience in Java, JEE, static analysis programming, stack trace reading, and source code inspection have an influence on applying PDM, as well as maintainers' previous experience with software development, software maintenance, and with the software that was used in the study.

The second knowledge question we wanted to answer (RQ2) concerned the acceptance of PDM by maintainers. Therefore, we used the TAM questionnaire (see Table 8) to evaluate the acceptance of PDM by the maintainers. Based on the TAM constructs, we answer the following questions:

RQ2. Would maintainers accept to use PDM?
    a.  How do maintainers perceive PDM regarding its ease of use?
    b.  How do maintainers perceive PDM regarding its usefulness?
    c.  Do maintainers intend to use PDM after experimenting it?

As TAM makes positive questions about the technology (see Table 8), we want to know the frequency in which maintainers agree with the questions. Additionally, we wanted to better understand the difficulties found by maintainers during PDM application. The frequency of certain difficulties found by maintainers might indicate their importance and improvement opportunities for PDM.

Table 9 and Table 10 respectively describe the set of independent and dependent variables together with their types and scales.

**Table 9: Independent variables**

| Type | Variables name and definition | Scale |
|------|------|------|
| Level of experience of maintainers | (L-Java) in Java<br>(L-JEE) in JEE<br>(L-STR) in stack trace reading<br>(L-SCI) in source code inspection<br>(L-SARP) in static analysis rule programming | 1 = No experience<br>2 = I studied in a classroom or in a book<br>3 = I actively practiced in a classroom project<br>4 = I used it in a project in industry<br>5 = I used it in several projects in industry |

| Time of experience of maintainers | (T-SD-I) in software development in industry (T-SM-I) in software maintenance in industry | Number of Years |
|---|---|---|

**Table 10: Dependent variables**

| Type | Variables name and definition | Scale |
|---|---|---|
| Percentage of maintainers | (P-CI-DP) that correctly identified all defect patterns (P-CD-DP) that correctly documented all defect patterns (P-Diff-DP) per reported difficulty found during defect pattern identification and documentation (P-CP-SARP) that correctly programmed the static analysis rule (P-Diff-SARP) per reported difficulty found during static analysis rule programming (P-CI-FA) that correctly identified all fixing alternatives (P-CD-FA) that correctly documented all fixing alternatives (P-Diff-FA) per reported difficulty found during identification and documentation of fixing alternatives (P-A-Q1) that agree or strongly agree in Q1 of TAM (P-A-Q2) that agree or strongly agree in Q2 of TAM (P-A-Q3) that agree or strongly agree in Q3 of TAM (P-A-Q4) that agree or strongly agree in Q4 of TAM (P-A-Q5) that agree or strongly agree in Q5 of TAM (P-A-Q6) that agree or strongly agree in Q6 of TAM (P-A-Q7) that agree or strongly agree in Q7 of TAM (P-A-Q8) that agree or strongly agree in Q8 of TAM (P-A-Q9) that agree or strongly agree in Q9 of TAM | Percentage (0% to 100%) |
| Number of defect patterns | (N-CD-D) correctly documented by a maintainer | Integer (0, 1 or 2) |

## 3.5. Experimental Procedures and Operation

The study started with the proper preparation of a laboratory with computers and Netbeans IDE for the subjects to be able to perform the tasks. As soon as subjects arrived, they received the consent form and the characterization form. After filling these forms, an introductory presentation of 30 minutes about PDM method was held, followed by a 20 minutes training on Task 1 activities.

This training included learning how to identify the data that should be extracted from the error logs, how to compare this data to identify similar failures, and how to compare similar failures in the source code to identify and document a defect pattern. The training of the Pilot group was slightly different from the one of groups A and B because the forms used for documenting failures and defect patterns were different. After training, they received a brief explanation about Task 1 and the materials of this task were distributed, i.e., the forms of task 1 together with the logs and the application source code. Participants had 40 minutes to perform Task 1, which consisted of extracting data of six failures from logs and identifying and documenting two defect patterns found in the application source code. At the end, they filled the follow-up form.

After performing Task 1, the same groups of maintainers performed Task 3. The Pilot group received a 10 minute break between Task 1 and Task 3, group A performed Task 1 and Task 3 in two different days, finally group B did not receive any interval between the tasks.

We started task 3 by distributing the forms of the task and the defect pattern documentation presented in Table 2. After that, we applied a training session of 20 minutes regarding Task 3. In this session, we showed how to identify false positives and how to document new defect fixing alternatives. Thereafter, the maintainers had 40 minutes to inspect 16 alerts of a defect pattern for classifying them into defects or false positives. The provided set of alerts contained 3 defects and 13 false positives that include two new fixing alternatives for the defect pattern. Finishing Task 3, maintainers filled the follow-up questionnaire. At the end of Task 3, we asked the maintainers to fill the TAM questionnaire.

We expected Task 2 to be more difficult and time-consuming than Task 3. Therefore, we were not able to apply Task 2 to group A and decided to apply Task 2 to group B of maintainers in a separate day. Group B performed this task directly using the source code abstract tree, and none of participants was able to complete the task. Hence, we decided to conduct new trials with groups C and D, using the SCPL DSL to better understand the feasibility and difficulties.

Hence, Task 2 was applied on a different time for groups C and D, which fosued solely on this specific task. Also, due to COVID-19 pandemic, it was applied for these groups in an on-line setup. We started Task 2 with an introductory presentation of 30 minutes about the PDM method. After that, a 20 minutes presentation of SCPL was held and the researchers helped students to prepare their computer's environment for using the tool. For operational reasons (group C had three hour classes once a week, while group D had two hour classes twice a week), Group C started to perform the rule programming part of the task right after the SCPL presentation while Group D had two days of interval before starting this part. The rule programming started by distributing the form of the task; then, the maintainers had 50 minutes to implement the rule. Finally, they filled the follow-up form.

## 4 Results

After executing the study, the materials were analyzed to determine the value of the variables to answer the stated questions. Some variables had to be determined by the researchers, such as the number of correctly identified, programmed or documented defect patterns. In these cases, one researcher inspected the forms and determined if the maintainers correctly identified, programmed or documented each item. The procedure of determining if a maintainer correctly identified a defect pattern or a fixing alternative included reviewing all fields in the respective form and determining if she captured the general idea of the pattern or fixing alternative. In another way, the procedure for deciding if documentation of a defect pattern or fixing alternative is correct, we searched for errors in the specific field of documentation. Finally, for deciding weather a custom static analysis rule was correctly programmed, the reasearchers inspected the code of the custom rule and the logs generated by its execution.

Additionally, qualitative data was analyzed to determine the most frequent difficulties of the maintainers. The qualitative analysis included open coding of the qualitative data using the constant comparative method (Seaman, 1999) and counting the most common codes.

The remainder of this section is organized by the tasks executed during the study, reporting on the effectiveness, the profile of most effective maintainers and their difficulties. The last section shows the analysis of PDM acceptance regarding the TAM questions stated for the whole method and not its isolated tasks. The answers and discussion of the research questions based on the analysis of the results follow in Section 5.

### 4.1. Task 1 – Failure Analysis and Defect Pattern Identification

Table 11 and Table 12 present the percentage and number of maintainers that correctly identified and documented none, one, or two defect patterns during task 1, respectively. Forty-eight maintainers completed task 1 by sending the task 1 form to the researchers. As task 1 had two defect patterns, the percentage of maintainers that were able to identify all defect patterns (P-CI-DP) ranged from 12% to 30%, whereas the percentage of maintainers that correctly documented them (P-CD-DP) ranged from 0% to 30%.

**Table 11: Percentage of maintainers that correctly identified defect patterns**

| Correctly Identified / Number of Defect Patterns | n | 0 | 1 | 2 |
|---|---|---|---|---|
| Pilot | 9 | 44% (4) | 44% (4) | 12% (1) |
| Group A | 23 | 52% (12) | 18% (4) | 30% (7) |
| Group B | 16 | 38% (6) | 50% (8) | 12% (2) |

**Table 12: Percentage of maintainers that correctly documented defect patterns**

| Correctly Documented / Number of Defect Patterns | n | 0 | 1 | 2 |
|---|---|---|---|---|
| Pilot | 9 | 89% (8) | 11% (1) | 0% (0) |
| Group A | 23 | 61% (14) | 9% (2) | 30% (7) |
| Group B | 16 | 44% (7) | 50% (8) | 6% (1) |

Regarding materials validation, we used two formats of defect pattern documentation. The Pilot (n=9) group used the form for defect pattern documentation while groups A and B (n=39) used pattern language. We found that individuals using the pattern language document more defect patterns correctly than those using the form (Wilcoxon-Mann-Whitney test on N-CD-DP, $p < 0.05$, one-sided). As we have found this difference, hereafter we discarded the Pilot group from our analyses, considering only Groups A and B (n=39), which used a pattern language for documenting defect patterns.

To understand differences in their profiles, we split maintainers into two groups, the ones that correctly identified and documented all patterns (success, n=8) and the complementary group (other, n=31). For each knowledge or experience variable (L-Java, L-JEE, L-STR, L-SCI, L-SARP, T-SD-I, T-SM-I), we compared their distribution using the Wilcoxon-Mann-Whitney test to evaluate if the success group had significantly higher values than the other one. The results of the tests are presented in Table 13. We found a significant difference ($p < 0.05$) for L-Java (level of experience with Java), L-STR (level of experience with stack trace reading), and L-SCI (level of experience with source code inspection) variables.

**Table 13: Results of Wilcoxon-Mann-Whitney tests between Success and Other groups in task 1 (n1=8, n2=31, one-tailed)**

| Variable | W | p |
| --- | --- | --- |
| L-Java | 169 | 0.026 |
| L-JEE | 128 | 0.441 |
| L-STR | 202 | 0.002 |
| L-SCI | 180 | 0.020 |
| L-SARP | 82 | 0.946 |
| T-SD-I | 162 | 0.077 |
| T-SM-I | 138 | 0.284 |

We analyzed the maintainers' challenges during task 1 based on the answers provided to an open question. Therefore, we open coded the qualitative data and counted the most common codes. Table 14 presents the results of this counting. As the groups of maintainers are different, we present the results separately per group. We can observe that the main difficulties of all groups involve somehow documenting and identifying defect patterns, i.e., in performing the task. Thereafter we also provide some examples of difficulties reported by the maintainers for Task 1.

**Table 14: Most frequent difficulties of maintainers in Task 1**

| Group of Maintainers | P-Diff-DP | Difficulty description (code) |
| --- | --- | --- |
| Group A | 5 of 23 (22%) | Identifying the patterns |
| | 5 of 23 (22%) | Documenting the patterns |
| | 5 of 23 (22%) | Identifying a solution for the defect pattern |
| | 4 of 23 (17%) | Lack of experience (Java, JEE and App Code) |
| Group B | 10 of 16 (63%) | Documenting the patterns |
| | 5 of 16 (31%) | Identifying the pattern |

Examples of difficulties reported by group A and the related codes are provided hereafter:

*"The main difficulty was confirming if the same exception types form a pattern in the source code." (Identifying the patterns)*

*"Representing the defect patterns in a generalized manner." (Documenting the patterns)*

*"To know the best way of fixing the code (with a try/catch or if/else)" (Identifying a solution for the defect pattern)*

*"No familiarity with the method, and the technology (servlets)" (Lack of experience)*

Examples of difficulties reported by group B and the related codes follow:

*"It was the first time I used this kind of form for documenting defect patterns. So, it took a while before it flows regularly" (Documenting the patterns)*

*"Recognizing patterns in different places of the source code and using the generic language to document the patterns." (Identifying the pattern and Documenting the patterns)*

## 4.2. Task 2 - Static Analysis Rule Programming

The groups that participated in this task were B, C, and D. Group B had 18 maintainers, but only eight remained until the end of the task sending the materials for evaluation. No maintainer was able to program the proposed static analysis rule correctly. I.e., the percentage of maintainers that correctly programmed the static analysis rule was 0%.

We also analyzed the maintainers' difficulties during Task 2 based on the answers provided to the related open follow-up question. Therefore, we open coded the qualitative data and counted the most common codes. Table 15 presents the most common codes found. We can observe that the time to perform the task was the main difficulty reported followed by the concepts involved and difficulties on implementation. Thereafter we also provide some examples of the difficulties reported by the maintainers of group B for Task 2.

**Table 15: Most frequent difficulties of maintainers of group B in Task 2**

| P-Diff-SARP | Difficulty (code) |
| --- | --- |
| 6 of 8, 75% | Time was not enough for performing the task |
| 4 of 8, 50% | Understand the concepts |
| 4 of 8, 50% | Difficulties on implementation |
| 3 of 8, 38% | Lack of experience |

Examples of difficulties reported by group B and the related codes are presented hereafter:

*"The time was short for understanding the topic, and this topic is complicated…" (Time was not enough for performing the task)*

*"Understand the methods, classes and how to use them to solve the problem." (Understand the concepts)*

*"Unfamiliarity with the topic, difficulty to implement the methods, even after understanding I did not know how to program." (Difficulties on implementation)*

*"Low knowledge regarding static analysis causing serious difficulties for completing the task."* (Lack of experience)

Motivated by these results, indicating that novice maintainers struggle to implement static analysis rules when directly using the source code AST, to further investigate this task, we conducted the additional trials with groups C (n=18) and D (n=10), using the SCPL DSL. Table 16 present the percentage and number of maintainers of groups C and D that correctly programmed the custom rule during Task 2. The percentage of maintainers that were able to programming the custom rule (P-CP-SARP) ranged from 10% to 33%. It is possible to observe that, using the DSL, in total 7 out of the 28 participants (25%) successfully completed the task.

**Table 16: Percentage of maintainers that correctly programmed the custom static analysis rule**

| Correctly Programmed Rule | n | Incorrect | Correct |
|---|---|---|---|
| Group C | 18 | 67% (12) | 33% (6) |
| Group D | 10 | 90% (9) | 10% (1) |

To understand the differences in the profile of the maintainers that correctly programmed the custom static analysis rule, we split maintainers into two groups, the ones that correctly programmed the rule (success, n=7) and the complementary group (other, n=21). It is noteworthy that most of maintainers that correctly programmed the rule are from group C. For each knowledge or experience variable (L-Java, L-JEE, L-STR, L-SCI, L-SARP, T-SD-I, T-SM-I), we compared their distribution using the Wilcoxon-Mann-Whitney test to evaluate if the success group had significantly higher values than the other one. The results of the tests are presented in Table 17. We did not find any significant difference between the two groups ($p < 0.05$) regarding higher experience values. However, we found a significant difference ($p = 0.005 < 0.05$) for a lower experience level of L-SARP (level of experience with static analysis rule programming) of the successful groups.

**Table 17: Results of Wilcoxon-Mann-Whitney tests between Success and Other groups in Task 2 (n1=7, n2=21, one-tailed)**

| Variable | W | p |
|---|---|---|
| L-Java | 65.5 | 0.673 |
| L-JEE | 57.0 | 0.833 |
| L-STR | 66.5 | 0.654 |
| L-SCI | 80.0 | 0.360 |
| L-SARP | 31.5 | 0.994 |
| T-SD-I | 57.5 | 0.821 |
| T-SM-I | 60.0 | 0.794 |

We also analyzed the difficulties of maintainers from groups C and D during Task 2 based on the answers provided in the related open-ended follow-up question. Therefore, we open coded the qualitative data and counted the most common codes. Table 18 presents the most common codes found. We can observe that the SCPL language syntax was the main reported difficulty, followed by static analysis rule programming and Java programming. Thereafter we also provide some examples of the difficulties reported by the maintainers of groups C and D for Task 2.

**Table 18: Most frequent difficulties of maintainers of groups C and D in Task 2**

| P-Diff-SARP | Difficulty (code) |
|---|---|
| 9 of 28, 32% | SCPL language syntax |
| 3 of 28, 10% | Static analysis rule programming |
| 3 of 28, 10% | Java programming |

Examples of difficulties reported by group C and D and the related codes are presented hereafter:

*"The biggest difficulty was describing the code format sought to find the possible error."* (SCPL language syntax)

*"As it is my first contact with this type of activity, I had a hard time knowing how to start."* (Static analysis rule programming)

*"I have a limited experience with Java programming. Especially with error treatment and SCPL, which I just learned how to use."* (Java programming, SCPL language Syntax)

*"Not being used to this type of analysis."* (Static analysis rule programming)

*"The biggest difficulty was to develop a java code that found patterns and avoided syntax errors."* (Java programming, SCPL language Syntax)

## 4.3. Task 3 - Rule Evaluation and Context Analysis

Table 19 and Table 20 present the percentage and number of maintainers that correctly identified and correctly documented none, one, or two fixing alternatives during task 3, respectively. 28 maintainers completed task 3 by sending the form to the researchers. As task 3 had two fixing alternatives, the percentage of maintainers that were able to identify all fixing alternatives (P-CI-FA) ranged from 0% to 65% within the groups A and B whereas the ones that correctly documented all fixing alternatives (P-CD-FA) ranged from 0% to 37.5%.

**Table 19: Percentage of maintainers that correctly identified fixing alternatives**

| Correctly Identified / Number of Fixing Alternatives | N | 0 | 1 | 2 |
|---|---|---|---|---|
| Group A | 20 | 5% (1) | 95% (19) | 0% (0) |
| Group B | 8 | 0% (0) | 37% (3) | 63% (5) |

**Table 20: Percentage of maintainers that correctly documented fixing alternatives**

| Correctly Documented / Number of Fixing Alternatives | N | 0 | 1 | 2 |
|---|---|---|---|---|
| Group A | 20 | 65% (13) | 35% (7) | 0% (0) |
| Group B | 8 | 25% (2) | 37.5% (3) | 37.5% (3) |

Again, to understand the differences in the profile of the maintainers that correctly identified and documented all fixing alternatives, we split maintainers into two groups, the ones that correctly identified and documented all fixing alternatives (success, n=3) and the complementary group (other, n=25). It is noteworthy that the three maintainers that correctly identified and documented all fixing alternatives are from group B. For each knowledge or experience variable (L-Java, L-JEE, L-STR, L-SCI, L-SARP, T-SD-I, T-SM-I), we compared their distribution using the Wilcoxon-Mann-Whitney test to evaluate if the success group had significantly higher values than the other one. The results of the tests are presented in Table 21. We did not find any significant difference between the two groups ($p < 0.05$).

**Table 21: Results of Wilcoxon-Mann-Whitney tests between Success and Other groups in Task 3 (n1=3, n2=25, one-tailed)**

| Variable | W | P |
| --- | --- | --- |
| L-Java | 31.5 | 0.712 |
| L-JEE | 47.5 | 0.213 |
| L-STR | 50.5 | 0.682 |
| L-SCI | 36 | 0.547 |
| L-SARP | 23.5 | 0.865 |
| T-SD-I | 18 | 0.942 |
| T-SM-I | 24 | 0.887 |

As done for the other tasks, we also analyzed the maintainers' difficulties during task 3 based on the answers provided to a related open follow-up question. Therefore, we open coded the qualitative data and counted the most common codes. Table 22 presents the results of this counting. As the groups of maintainers are different, we present the results separately per groups. Thereafter, we also present some examples of the reported difficulties of maintainers.

**Table 22: Most frequent difficulties of maintainers on task 3**

| Group of Maintainers | P-Diff-FA | Difficulty |
| --- | --- | --- |
| Group A | 6 of 20, 30% | Understand the app source code |
| Group B | 3 of 8, 30% | Lack of experience with the task |

Example description of the difficulty reported by a participant of Group A:

*"Sometimes the scope of the method can be really big, so inspecting every scope to check validation functions can become confusing."* [6]  (Understand the app source code)

Example description of the difficulty reported by a participant of Group B:

---

[6] One point of inspection (file PrincipalTermo.java line 378) is inside a long method (more than 400 lines of code). Although inspecting this method was a challenge for maintainers, a simple static analysis rule (see Figure 4) produced by PDM can correctly classify this line as having no defect.

*"Lack of practice in activities like this."* (Lack of experience with the task)

### 4.4. TAM – Technology Acceptance Model

At all, 27 maintainers from groups A and B filled the TAM questionnaire. Maintainers of groups C and D did not answer the TAM questionnaire, as the questionnaire focused on PDM and groups C and D concerned an additional evaluation focused only on the static analysis rule programming task. Table 23 presents the percentage of the maintainers that agree or strongly agree with each question of TAM questionnaire (See Table 8). We can observe that most of maintainers found PDM useful, but not easy to use, and that they do not intend to use PDM in their work.

**Table 23: Percentage of maintainers that strongly agree or agree with TAM questions regarding PDM method.**

| TAM Dimension | Question | Percentage of Maintainers that Strongly Agree or Agree |
|---|---|---|
| Usefulness | Q1 | 77% |
| | Q2 | 74% |
| | Q3 | 81% |
| | Q4 | 70% |
| Ease of use | Q5 | 26% |
| | Q6 | 30% |
| | Q7 | 34% |
| | Q8 | 34% |
| Intention to use | Q9 | 15% |

Further discussions of the results, answering the research questions, follow in the next section.

## 5 Discussion

In this section, we answer and discuss our research questions, presenting the main related findings and insights.

*RQ1. What are the challenges faced by maintainers while applying PDM for creating custom static analysis rules?*
    *a. What are the challenges faced by maintainers while identifying and documenting defect patterns?*

Regarding the effectiveness of maintainers applying PDM, only 12% of the maintainers in group A and 30% in group B identified and documented all defect patterns. These results showed that after 20 minutes of training in task 1, few maintainers could effectively apply it. This fact could indicate that more training would be needed or that the application of task 1 of PDM needs to be facilitated in some way.

We collected and analyzed the challenges found by maintainers during PDM application. During Task 1, the documentation format hindered the pilot group in documenting defect patterns. We could notice this problem by observing maintainer's comments on difficulties in this task and a statistically significant difference between pilot group and the other groups. After changing the documentation format, groups A and B performed better than the pilot group in documenting defect patterns. However, groups A and B also complained about difficulties in identifying and documenting defect patterns, showing that this task could be indeed tricky.

The maintainers that correctly identified and documented defect patterns had superior experience in Java, stack trace reading, and source code inspection. Thus, this type of knowledge possibly influences in performing task 1. The levels of experience of maintainers that successfully completed task 1 were in its majority industrial, while other maintainers had only academic experience. This fact might indicate that an appropriate profile for performing task 1 is a maintainer with industrial experience in bug fixing using the technology of the software in which PDM will be applied. Indeed, industrial bug fixing activity typically involves stack trace reading and source code inspection. Hence, the training of maintainers to perform Task 1 should include those skills.

### b. *What are the challenges faced by maintainers while programming a custom static analysis rule?*

When group B conducted Task 2, no maintainer was able to program the proposed static analysis rule correctly. The main complaint of maintainers on this task was the limited time for programming (50 min), the difficulty to comprehend the concepts involved (Abstract Syntax Tree and Visitor Pattern), and difficulties for programming the rule using those concepts and SonarQube. This might indicate that programming static analysis rules directly using the AST is a particular task and that it might be difficult to find a professional that is already skillful at it. In the case of training professionals in AST-based static analysis rule programming, a single training session with 20 minutes of presentation and 50 minutes of exercises will not suffice.

Regarding the effectiveness of programming a custom static analysis rule using SCPL, 33% of the maintainers in group C and 10% in group D correctly performed the task. Most of the maintainers that correctly programmed the rule were from Group C. This group had two days of interval between the training session on SCPL and the task performing session of programming the rule, while Group D had no interval between those sessions. We hypothesize that maintainers could have used this time to study the technology and practice before performing the task. Hence, the time factor may have influence in the understanding of the technology used for programming the rule and consequently in maintainers' performance on using it.

Surprisingly, the maintainers that correctly programmed the static analysis rule of Task 2 reported having less experience in programming static analysis rules than the others. A possible explanation would be that less experience in the task could have led to a greater interest in understanding and using the technology. However, further investigation is required to better understand this result. No other significant difference in any knowledge and experience variables to other maintainers were found.

The main complaints of maintainers of groups C and D regarding Task 2 were about the SCPL language syntax, static analysis programming, and Java programming. Those complaints are comprehensible since the maintainers had no previous experience with SCPL and few of them had previous experience with Java and static analysis rule programming. In spite of those difficulties, some of the maintainers correctly completed the task. This indicates that effectiveness in programming custom rules with SCPL could be more influenced by the time for understanding and internalizing the technology than by previous knowledge and experience. In this way, the training of maintainers to perform static analysis rule programming using a DSL should focus on the DSL syntax and include plenty of examples and time to practice.

  *c. What are the challenges faced by maintainers while identifying and documenting fixing alternatives present in false positives of a rule?*

  Similar to Task 1, few maintainers were able to correctly identify and document fixing alternatives during Task 3. No maintainer of group A and 63% of the maintainers of group B correctly identified all fixing alternatives while no maintainer of group A and 37.5% of group B correctly documented them. The maintainers that correctly identified and documented all fixing alternatives had no significant difference in any knowledge and experience variables to other maintainers. As stated before, group B had previous experience with the software to which PDM was applied by using it in the course assignment. This fact might indicate that previous experience with the software being maintained positively affects the performance of identifying fixing alternatives.

  The main complaints of maintainers regarding Task 3 for the group A were related to understanding the application source code, and for group B to the lack of experience on the task. Group A's complaint reinforces the insight that experience with the software affects the performance of Task 3. When the maintainer has this experience, as in the case of group B, the main complaint was not the software but the experience with the task. Hence, the training of maintainers to perform Task 3 should include the comprehension of the subject software source code and its structures.

  *RQ2. Would maintainers accept to use PDM?*
    *a. How do maintainers perceive PDM regarding its ease of use?*
    *b. How do maintainers perceive PDM regarding its usefulness?*
    *c. Do maintainers intend to use PDM after experimenting it?*

  The TAM questionnaire was used to assess the PDM acceptance by maintainers. Table 23 shows that most maintainers perceive PDM as useful but not easy to use. This perception may have an influence on the intention to use PDM since few maintainers answered agreeing with this question in the TAM questionnaire. These perceptions show that PDM application should be facilitated for improving its acceptance by maintainers.

# 6 Threats to Validity

  **Internal Validity.** During the study, some maintainers did not perform all tasks. As an uncontrollable condition of mortality, some maintainers left the experimentation session before completing tasks 2 and 3, especially in group B. This group had classes at night, and as the end of the class come close some of them naturally left the class. This condition made the number of subjects vary significantly from task 1 (n=39) to task 3 (n=28). Hence, the results of Task 3 might have been affected by this variation. Furthermore, to reduce researcher bias, the coding performed by one of the researchers during the qualitative analysis was peer-reviewed by a second researcher.

**Construct Validity.** The task selection, order of application, and time to complete the tasks could have affected the results of the study. We have selected to perform Task 1 and Task 3 in groups A and B, while we applied Task 2 in groups B, C and D. This task to group assignment was made considering the constrained time in the disciplines of the groups. As PDM steps interact, the changing in order or avoiding steps might affect the effectiveness in applying PDM and the perception of maintainers about the method. Furthermore, the time of training for performing each task was also constrained by the time available in classes of the groups and might have affected the results of the study. Finally, Task 2 was applied in an on-line setup for groups C and D due to COVID-19 pandemic. In on-line environments, it is not possible to avoid that students interact with each other during the experimentation. This factor may have had influence in the results of Task 2.

**Conclusion Validity.** Our purpose was to conduct an observational study to evaluate if other maintainers would be able to apply the PDM steps. Given our limited sample size, we had no further aims regarding conclusion validity. Indeed, while some maintainers successfully completed the steps, their number was not enough to make claims about their characteristics. It is noteworthy that, although we used a large group of students in our research (n=82), they were split into five different groups, and only a few of them correctly completed each task. Hence, while we analyzed the characteristics of those groups against their complementary groups, the confidence in the results is affected by our sample size.

**External Validity**. Our observational study considered a specific setting (e.g., software, technology, students). Hence, external validity is limited. Furthermore, as common with empirical studies conducted with students, the results concern novice maintainers and their characteristics, not being generalizable. Additionally, the changes in programming language and other variables from the industrial study (Mendonça et al., 2018) to our observational study, do not allow us to compare the actual usefulness of PDM with the perceived one.

# 7 Conclusion

Although custom static analysis rules are perceived as useful (Tymchuk et al., 2018), they are still not widely adopted in practice (Beller et al., 2016; Christakis & Bird, 2016). In this study, we performed an observational study on the challenges of creating custom static analysis rules in the context of a systematic method to create those rules, called Pattern-Driven Maintenance (PDM) (Mendonça et al., 2018).

We evaluated maintainers' effectiveness in applying PDM and their acceptance of the method. In this way, we observed novice maintainers applying PDM steps split into three tasks, i.e., failure analysis and defect pattern identification (Task 1), static analysis rule programming (Task 2), and rule evaluation and context analysis (Task 3). Regarding Task 2, we investigated AST-based and DSL-based rule programming.

The maintainers had difficulties during PDM steps application, and few of them correctly completed the tasks. The difficulties found included the defect pattern documentation format, which was changed during the study, the identification of defect patterns and their fixing alternatives, the static analysis rule programming, and the understanding of the software source code. Maintainers also answered a TAM questionnaire about PDM acceptance. Most of them found PDM useful but not easy to apply and do not intend to use PDM at work. However, the perceived ease of use of PDM could have been hindered by the conditions of limited time of an observational study, thus affecting the intention of use.

We analyzed the profile of maintainers that correctly completed the tasks. We found that the ones that correctly completed Task 1 had superior experience in the software programming language (Java), stack trace reading, and source code inspection; while in Task 3 the most effective maintainers had previous experience with the software to which PDM was applied. In Task 2, we initially investigated AST-based rule programming. However, even with a short training session, none of them was able to correctly implement the rule. Hence, we concluded that expecting developers which are not rule experts to develop defect rules directly using the AST is unrealistic. Therefore, we conducted additional trials to investigate DSL-based rule programming, using SCPL (Crispe & Mendonça, 2021). In these trials, we observed that some (25%) of the maintainers were able to correctly implement the rule and that time to learn the DSL influenced the effectiveness of rule programming.

The results strengthen our confidence that PDM can help maintainers in producing custom static analysis rules for locating defects. However, a proper selection and training of maintainers is needed to apply PDM effectively. Also, using a higher level of abstraction, such as a DSL, can ease static analysis rule programming for novice maintainers.

In future work, we intend to evaluate PDM acceptance by experienced maintainers and using other technologies for rule programming. We do believe that experienced maintainers could have different results on applying PDM than novice ones. Their feedback can help to further improve the PDM method and to better evaluate PDM acceptance.

## Acknowledgments

This work was partially supported by the CNPq grant 141345/2015-2.

## Declarations

### Funding
This work was partially supported by the CNPq grant 141345/2015-2.

### Conflicts of interests
The authors have no relevant financial or non-financial interests to disclose.